\documentclass[12pt,english]{article}
\usepackage[breaklinks,hypertexnames=false]{hyperref}
\hypersetup{colorlinks=true,linkcolor=blue,urlcolor=blue,citecolor=blue}
\usepackage{amsthm}
\usepackage{amsmath}
\usepackage{amssymb}
\usepackage{esint}
\usepackage[authoryear]{natbib}

\usepackage{subcaption} 

\usepackage{appendix}
\usepackage[english]{babel}
\usepackage{enumitem}
\usepackage{booktabs}
\usepackage{multirow}
\usepackage{natbib}
\usepackage{textcomp,mathcomp}
\usepackage{adjustbox}
\usepackage{dcolumn}
\usepackage{booktabs}
\usepackage{mathabx} 

\makeatletter
\usepackage{graphicx}
\usepackage{lscape}
\usepackage{rotating}
\usepackage[english]{babel}
\usepackage{latexsym}	
\usepackage{bbm}
\usepackage{tikz}
\usetikzlibrary{arrows,shapes,trees,positioning}
\usepackage{seqsplit}

\hypersetup{
	colorlinks=true,
	citecolor=blue
}

\makeatother

\newtheorem{assumption}{Assumption}

\newtheorem{example}{Example}

\newtheorem{remark}{Remark}

\newcommand{\defeq}{\vcentcolon=}

\usepackage{todonotes}
\usepackage{tikz}

\newcommand{\wheelchart}[3]{
	\pgfmathsetmacro{\totalnum}{0}
	\foreach \value/\colour in {#1} {
		\pgfmathparse{\value+\totalnum}
		\global\let\totalnum=\pgfmathresult
	}
	
	\pgfmathsetmacro{\wheelwidth}{(#3)-(#2)}
	\pgfmathsetmacro{\midradius}{(#3+#2)/2}
	
	\begin{scope}[rotate=90]
		\pgfmathsetmacro{\cumnum}{0}
		\foreach \value/\colour in {#1} {
			\pgfmathsetmacro{\newcumnum}{\cumnum + \value/\totalnum*360}
			
			\draw[fill=\colour] (-\cumnum:#2) arc (-\cumnum:-\newcumnum:#2)--(-\newcumnum:#3) arc (-\newcumnum:-\cumnum:#3)--cycle;
			
			\global\let\cumnum=\newcumnum
		}
	\end{scope}
}

\input{tcilatex}
\usepackage{babel}

\begin{document}
	
		\title{Estimating Heterogeneous Effects:\\Applications to Labor Economics\thanks{We thank Evan Rose for helpful comments. The views expressed in this paper are those of the authors and do not necessarily reflect the position of the Banco de España or the Eurosystem.}}
	\author{St\'{e}phane Bonhomme \\ University of Chicago \and Angela Denis \\ Banco de España }
	\date{\today
	}
	\maketitle
	\vskip 1cm
	
	\begin{abstract}
		
		A growing number of applications involve settings where, in order to infer heterogeneous effects, a researcher compares various units. Examples of research designs include children moving between different neighborhoods, workers moving between firms, patients migrating from one city to another, and banks offering loans to different firms. We present a unified framework for these settings, based on a linear model with normal random coefficients and normal errors. Using the model, we discuss how to recover the mean and dispersion of effects, other features of their distribution, and to construct predictors of the effects. We provide moment conditions on the model's parameters, and outline various estimation strategies. A main objective of the paper is to clarify some of the underlying assumptions by highlighting their economic content, and to discuss and inform some of the key practical choices.

		\bigskip
		
		
	\end{abstract}
	
	\clearpage
	
	\global\long\def\ind{\mathbb{1}}
	\global\long\def\d{\mathrm{d}}
	\global\long\def\t{\intercal}
	\global\long\def\RR{\mathbb{R}}
	\global\long\def\defeq{:=}


	\section{Introduction\label{secIntro}}

Documenting heterogeneity across individuals, firms, or space, has become a central theme in applied economics. In earlier research, heterogeneous parameters used to be treated as a nuisance, and ``differenced out'' by means of fixed-effects regressions. Increasingly, however, estimating and studying heterogeneous effects has become the main goal of the analysis.

This focus on heterogeneity is enabled by the availability of richer data sets. A leading example is given by administrative data sets that feature many units (such as firms, workers, or neighborhoods), and at the same time provide information about each unit (such as multiple workers in a firm, multiple time periods on a worker, or multiple individuals in a neighborhood).

Settings with heterogeneous parameters are often studied using panel data techniques. However, traditional panel data methods treat units, such as firms or neighborhoods, independently of each other. A growing number of applications instead involve settings where, in order to infer heterogeneous effects, the researcher specifies a research design that compares various units. For example, to estimate firm or neighborhood effects, researchers exploit workers who move between firms or children who move to a new neighborhood, respectively.


We will refer to three leading examples as illustrations. In the first one, \citet{kline2022systemic} send multiple job applications to several large firms, and estimate firm-specific call-back rates as a function of applicants' characteristics. Documenting differences across firms in call-back rates allows the researchers to study hiring discrimination at the firm level, hence complementing the literature using r\'esum\'e correspondence experiments (e.g., \citealp{bertrand2017field}) by providing firm-level estimates. In this application, discrimination parameters can be estimated independently, firm by firm. Hence, this setting is akin to a traditional panel data or grouped data setting.

In the second example that we analyze, \citet{chetty2018impacts2} study how the income at adulthood of children depends on the place where they grew up. To estimate the effects of neighborhoods on income, the researchers exploit mobility of families \emph{across} neighborhoods. Hence, the effect of neighborhood $j$ on income at adulthood is constructed by comparing the incomes of individuals who grew up in various neighborhoods $j'$. This setting has been studied by a large subsequent literature, including \citet{chetty2018impacts}, \citet{laliberte2021long}, \citet{bergman2019creating}, and \citet{aloni2023one}.   

Our third leading example is the so-called AKM regression framework for matched employer-employee data introduced by \citet{abowd1999high}. The researchers' goal is to estimate how worker and firm effects contribute to wage dispersion. This setting similarly features comparisons between multiple units, since the differences in wage premia offered by two firms $j$ and $j'$ is informed by workers moving between $j$ and $j'$. The AKM approach has become central to the study of workers and firms, see among others \citet{card2013workplace}, \citet{card2018firms}, and \citet{song2019firming}. Moreover, the methodology in this example has been used to study other questions, such as differences in health care utilization across space inferred from patients moving between cities (\citealp{finkelstein2016sources}), or the impacts of banks and firms on credit growth inferred from banks' loans to multiple firms (\citealp{amiti2018much}).

In these applications, the data sets are complex and the models are high-dimensional. Practitioners need to make a large number of choices for modeling, practical specification, and estimation. Since we have not seen any survey on the econometric analysis of these settings as of yet,\footnote{\citet{abowd2008econometric} and \citet{bonhomme2020econometric} survey methods for bipartite networks and matched employer-employee data.} we have decided to write a paper on the topic. Our main goal is to lay out a framework to analyze these settings, and to clarify some of the underlying assumptions and key practical choices. While doing so, we will highlight the economic content behind the main assumptions.

The model that underlies most applications to these settings is a linear normal random coefficients (RC) model. The ``random coefficients'' refer to the heterogeneous effects that are the focus of the analysis, such as the effects of neighborhoods, or the worker and firm effects. Those coefficients are associated with specific covariates: in \citet{chetty2018impacts2} the effect of a neighborhood is simply the coefficient of the exposure to that neighborhood (i.e., of how long the family stayed in that neighborhood), whereas in \citet{abowd1999high} the effect of firm $j$ is the coefficient of the $j$-th firm indicator. In both cases, the model involves a very large number of such covariates (e.g., many thousands of firm and worker indicators or neighborhood exposures).


%
%

The primitive parameters of the RC model are the means and variances of the coefficients (e.g., the neighborhood effects, or the worker and firm effects), as well as the variance of the errors. All of these parameters are potentially functions of all the covariates. They satisfy first and second moment conditions that we present. These moment conditions, which remain valid absent normality, build on moment conditions previously derived for panel data settings (e.g., \citealp{chamberlain1992efficiency}, \citealp{arellano2012identifying}). 

While means and variances are useful to answer substantive questions, such as how much dispersion in outcomes is explained by neighborhood, worker, or firm heterogeneity, other important questions require additional information. As we describe, higher-order moments (such as skewness or kurtosis), nonlinear moments, as well as marginal and multivariate distributions, can all be inferred under the normality assumption of the RC model. For example, researchers may be interested in the distribution of neighborhood effects across space, or in the densities of worker and firm effects. In addition, the normal RC model can be used to construct optimal predictors of the effects. Such predictors or ``forecasts'' are also of considerable interest in various literatures outside our three leading examples, including in the work on teacher quality (e.g., \citealp{kane2008estimating}, \citealp{chetty2014measuring}).


Taking the RC model to data, however, is challenging in the settings that we study, given the large number of covariates (such as neighborhood exposures, or worker and firm effects) and the complex forms of dependence implied by the model. We show that commonly used specifications effectively impose conditional independence assumptions that may be economically restrictive. For example, common strategies in the analysis of neighborhood effects require location choice to be independent of neighborhood heterogeneity conditional on a set of neighborhood-specific covariates. Estimating flexible models of means, covariances, and distributions in these settings is an important yet still relatively unexplored research area.

Lastly, while the normal RC model provides a unified, self-contained framework to estimate heterogeneous parameters, normality and linearity may both be restrictive. In the last part of the paper, we briefly explore how these assumptions could be relaxed.

\paragraph{Relation to the literature.}

The framework we present builds on a vast methodological literature in statistics and econometrics. This includes the statistical literature on mixed models (e.g., \citealp{jiang2007linear}, \citealp{mcculloch2004generalized}), the literature on random coefficients models and correlated random-effects approaches in panel data (e.g., \citealp{chamberlain1992efficiency}, \citealp{arellano2012identifying}) and related panel data work based on decision-theoretic approaches (\citealp{chamberlain2009decision}, \citealp{chamberlain2016fixed}), as well as the empirical Bayes literature (e.g., \citealp{efron2012large}) and the Bayesian and frequentist interpretations of best linear unbiased predictors (BLUP, e.g., \citealp{robinson1991blup}).

	\section{Regressions with fixed effects\label{secExamples}}

It is common in empirical work to focus on the relationship between some (typically scalar) outcomes $y_i$ and some covariates $x_i$ and $z_i$, and to specify a linear regression model of the form
\begin{equation}
	y_i=x_i'\beta+z_i'\eta+u_i,\quad i=1,...,n.\label{eq_micro}
\end{equation}

 We assume that the researcher has access to multiple observations $i$ about some economic units $j$. The units may be firms, workers, or neighborhoods, depending on the application. Our focus is on settings where $z_i$ contains unit-specific variables, as well as interactions of unit-specific variables with other covariates. Let $p$ denote the number of covariates in $z_i$. Hence, the coefficients $\eta_j$, $j=1,...,p$, are unit-specific ``fixed effects''.\footnote{We will refer to the $\eta_j$'s as ``fixed effects'', in line with the usual terminology in applied economics. In contrast, in the statistical literature on mixed models $\beta$ are often referred to as ``fixed effects'', and $\eta$ as ``random effects''. See, e.g., the monograph by \citet{jiang2007linear}.} Let $q$ denote the number of covariates in $x_i$. We will focus on settings where $p$ is large, and, depending on the application of interest, $q$ may be small or large. 

Estimating high-dimensional regressions as in (\ref{eq_micro}) is made possible by the availability of increasingly large and detailed data sets, and by the development of powerful computational methods and software. This enables researchers to ``zoom in'' on the effects of particular units in the sample.  

Under the assumption that $u_i$ are uncorrelated with $x_i$ and $z_i$ in (\ref{eq_micro}), researchers typically estimate $\beta$ and $\eta$ using OLS. This regression delivers parameter estimates $\widehat{\beta}$ and $\widehat{\eta}$. The researcher's goal is then to use the estimates $\widehat{\eta}_1,...,\widehat{\eta}_p$ to learn about the true effects $\eta_1,...,\eta_p$.

We now describe three examples where this setup arises.

\begin{example}{\textbf{(Firm-specific discrimination)}}\label{Ex_1} \citet{kline2022systemic} construct firm-specific measures of racial discrimination in hiring. They send various fictitious job applications with randomized characteristics to some of the largest employers in the US. They then measure call-back rates by race, for every firm in the sample. To map this setting to the current setup, denote job applications as $i$ and firms as $j$. The outcome $y_i$ is whether the firm calls back the applicant, and $z_i$ contains the firm's indicator and the interaction of the firm's indicator with the race of applicant (white or black). There are no additional covariates $x_i$. The researchers are interested in the coefficient in $\eta_j$ of the interaction between firm $j$'s indicator with the race of the applicant, which they interpret as reflecting firm $j$'s discrimination in hiring.

In this particular setting, the regression coefficient $\widehat{\eta}_j$ is constructed using the observations from firm $j$ only, not other firms' observations. The data structure is analogous to panel (or grouped) data, and indeed, in this application, model (\ref{eq_micro}) can be interpreted as a grouped data model with group-specific intercepts and slopes, where the slope coefficients are associated with the applicant's race. 



\end{example}

Increasingly, researchers estimate regressions where learning about $\eta_j$ requires comparing various units. In that case, the estimate $\widehat{\eta}_j$ for unit $j$ is constructed using the observations from several (and possibly a large number of) other units $j'$. Our next two examples are in this vein.

\begin{example}{\textbf{(Neighborhood effects and intergenerational mobility)}}\label{Ex_2} \citet{chetty2018impacts2} estimate the effects of neighborhoods in the US (such as counties or commuting zones) on income at adulthood. In the setting they propose, $i$ are children, the outcome $y_i$ is the income of a child at age 26 (specifically, her rank in the overall income distribution at age 26), $z_i$ contains the time spent by the child in every neighborhood when young as well as its interaction with parental income, and $x_i$ contains origin times destination indicators, also interacted with parental income, as well as interactions between children's cohort indicators and parental income. The times spent by the child in every neighborhood, which the authors refer to as the ``exposures'' to the neighborhoods, are key covariates in the model. Note that, in this case, both $z_i$ and $x_i$ are high-dimensional. 

The researchers are interested in neighborhood $j$'s effect on adult outcomes at some particular level of parental income $\overline{p}$. This neighborhood-specific parameter is a linear combination of the $\eta_j$ parameters that pertain to neighborhood $j$. The exposure-time research design, which builds on \citet{chetty2018impacts}, implies that the estimate of neighborhood $j$'s effect depends on the outcome data on other neighborhoods $j'$. The authors rely on this design to estimate the causal effect of neighborhoods, under the assumption that the age at which children move across neighborhoods does not directly affect adult outcomes.


\end{example}

\begin{example}{\textbf{(Firm and worker effects in wage determination)}}\label{Ex_3} \citet{abowd1999high} study how worker and firm heterogeneity contribute to wage dispersion. In the setup they propose, $y_i$ are log wages in a given period, $x_i$ includes age and time indicators as well as other demographics, and $z_i$ includes worker indicators and firm indicators. The worker and firm indicators are key covariates in the model. The researchers are interested in the coefficients $\eta_j$ associated with workers and firms. A first question of interest is how dispersed are those worker and firm effects? By answering this question, the AKM model sheds light into how much of wage dispersion can be attributed to workers earning different wages irrespective of where they work, versus firms paying similar workers differently. A second question is how correlated are worker and firm effects? By answering this second question, the model sheds light on the nature of sorting patterns and how sorting contributes to wage dispersion. 

Recovering worker and firm effects requires exploiting movements between firms. Intuitively, if workers remain in the same firms over time it is not possible to tell whether a high wage reflects a high worker effect or a high firm effect. Hence, the estimates of the effects depend on the network of employment relationships between workers and firms. Indeed, identification of the effects requires the network to be connected (\citealp{abowd2002computing}). 


\end{example}

\section{Quantities of interest and noisy estimates\label{secAnatomy}}

In the three applications that we have outlined, the researcher's goal is to use the estimates $\widehat{\eta}_1,...,\widehat{\eta}_p$ to learn about the true effects $\eta_1,...,\eta_p$. We now provide examples of quantities of interest taken from the literature.

Throughout the paper we will treat $\eta_1,...,\eta_p$ as random. Our goal will be to estimate features of the joint distribution of $\eta_1,...,\eta_p$, and construct predictors of those effects, without imposing \textit{a priori} that the $\eta_j$'s are independent of each other or independent of the $x_i$'s and $z_i$'s. When the conditional distribution of $\eta_1,...,\eta_p$ given $x_1,...,x_n,z_1,...,z_n$ is left unrestricted, proceeding in this way does not materially differ from a setup where the $\eta_j$'s are treated as fixed parameters (i.e., as ``fixed effects''). The model we will present in Section \ref{secModel} will impose restrictions on this conditional distribution, however. 


\subsection{Quantities of interest\label{subsec_interest}}

Researchers may have various objectives. A first goal may be to estimate some moments of $\eta_j$. For simplicity we will focus on the case where $\eta_j$ is scalar, although the expressions below are easily adapted to the case of vector-valued $\eta_j$'s.

Moments that are expectations of linear combinations of the $\eta_j$'s can be written as
\begin{equation}
	m_c=\mathbb{E}\left[c'\eta\right],\label{eq_lin_comb}
\end{equation}
where $c$ is a $p\times 1$ vector. An example is the mean of the $\eta_j$'s, 
$$\mathbb{E}\left[\frac{1}{p}\sum_{j=1}^p\eta_j\right].$$ 

Consider the coefficient in the linear regression of $\eta_j$ on some covariates $W_j$. For example, one may be interested in regressing firm effects on firm size or industry indicators in Example \ref{Ex_3}, or in regressing neighborhood effects on average income in the neighborhood in Example \ref{Ex_2}. The regression coefficient can be written as $\left(\mathbb{E}\left[W_jW_j'\right]\right)^{-1}\mathbb{E}\left[W_j\eta_j\right]$, which takes the form (\ref{eq_lin_comb}) for $c=\left(\mathbb{E}\left[W_jW_j'\right]\right)^{-1}W_j$.


Moments that are quadratic in $\eta$ can be written as
\begin{equation}
	v_Q=\mathbb{E}\left[\eta'Q\eta\right],\label{eq_quad_form}
\end{equation}
where $Q$ is a $p\times p$ matrix. For example, the variance of the $\eta_j$'s, $$\limfunc{Var}(\eta_j)=\mathbb{E}\left[\frac{1}{p}\sum_{j=1}^p\left(\eta_j-\frac{1}{p}\sum_{j'=1}^p\eta_{j'}\right)^2\right],$$ can be written in this form for a suitable matrix $Q$. In Example \ref{Ex_3}, interest often centers on the variances of worker and firm effects and on the covariance between worker and firm effects, all of which can be written as (\ref{eq_quad_form}). 

Alternatively, one may be interested in the coefficient in the linear regression of some scalar variable $W_j$ on $\eta_j$. For example, in Example \ref{Ex_3} one may be interested in regressing promotion opportunities in a firm on the firm effects. The regression coefficient is $\left(\mathbb{E}\left[\eta_j^2\right]\right)^{-1}\mathbb{E}\left[\eta_jW_j\right]$, which can be written as the ratio between some $m_c$ in (\ref{eq_lin_comb}) and some $v_Q$ in (\ref{eq_quad_form}), for suitable $c$ vector and $Q$ matrix.  

 


More general, nonlinear moments can be written as
\begin{equation}
	w_H=\mathbb{E}\left[H(\eta)\right],\label{eq_nonlin}
\end{equation}
for some function $H:\mathbb{R}^p\rightarrow \mathbb{R}$. For example, one may be interested in the skewness or kurtosis of the $\eta_j$'s, which can be written as ratios of quantities of the form (\ref{eq_nonlin}) for suitable $H$ functions. 

Learning about distributions of effects, beyond their means and variances, is important in all the examples we have mentioned. In Example \ref{Ex_1}, \citet{kline2022systemic} report estimates of the distribution of racial discrimination in hiring across firms. In Example \ref{Ex_2}, researchers are often interested in documenting the distribution of neighborhood effects. In Example \ref{Ex_3}, an increase in the variance of firm effects, say, has different implications for inequality whether it comes from a deepening of the left tail, an expansion of the right tail, or a symmetric increase in spread. 

The weighted cumulative distribution function, for some weights $\omega_j$ that sum up to one, can be written as the following nonlinear moment    
\begin{equation}
	F_{\omega}(a)=\mathbb{E}\left[\sum_{j=1}^p\omega_j\boldsymbol{1}\left\{\eta_j\leq a\right\}\right].\label{eq_cdf}
\end{equation}
One may also be interested in the (weighted) density of the $\eta_j$'s, $f_{\omega}(a)=\frac{\partial F_{\omega}(a)}{\partial a}$. Bivariate counterparts to $F_{\omega}(a)$ and $f_{\omega}(a)$, which reflect the bivariate distribution of workers and firms, are of interest in Example \ref{Ex_3} in order to document sorting patterns along the worker and firm distributions.


Another common goal in applications is to construct predictors of the $\eta_j$'s. The optimal predictor that minimizes the expected sum of squared errors is a set of functions $\varphi_1,...,\varphi_p$ that solves 
\begin{equation}\limfunc{min}_{\varphi_1,...,\varphi_p}\, \mathbb{E}\left[\sum_{j=1}^p\left(\eta_j-\varphi_j(y_1,...,y_n,x_1,...,x_n,z_1,...,z_n)\right)^2\right],\label{eq_MSE}\end{equation}
the solution of which is the set of conditional means
\begin{equation}
	\varphi_j(y_1,...,y_n,x_1,...,x_n,z_1,...,z_n)=\mathbb{E}\left[\eta_j\,|\, y_1,...,y_n,x_1,...,x_n,z_1,...,z_n\right],\quad j=1,...,p.\label{eq_post_means}
\end{equation}
In Example \ref{Ex_2}, \citet{chetty2018impacts} and \citet{chetty2018impacts2} construct predictors of neighborhood-specific income effects. \citet{bergman2019creating} use effect predictors to select the top census tracts for income mobility.\footnote{See \citet{gu2023invidious} for a broad account of ranking problems and selection of groups of units based on noisy estimates.} A key input to answering these questions is the set of conditional means given by (\ref{eq_post_means}).




\subsection{Noisy estimates}

The data is not directly informative about the $\eta_j$'s, but delivers estimates
\begin{equation}
	\widehat{\eta}_j=\eta_j+v_j, \quad j=1,...,p,\label{eq_decomp}
\end{equation}
where $v_j=\widehat{\eta}_j-\eta_j$ reflects estimation noise. In many applications involving large data sets (large $n$) and many unit-specific parameters (large $p$), the noise $v_j$ is substantial enough to be of practical concern. Inferring $\eta_j$ from $\widehat{\eta}_j$ then requires solving a filtering (or ``de-noising'') problem.

It is important to note that directly using the estimates $\widehat{\eta}_j$ in place of $\eta_j$ may be misleading. The noise $v_j$ in (\ref{eq_decomp}) reflects the presence of a form of measurement error. When the quantity of interest is nonlinear in $\eta_j$, such as a variance, a higher-order moment, a cumulative distribution function, or a density, the presence of measurement error often leads to unreliable, biased estimates of the quantities of interest.

As an example, suppose the researcher is interested in the variance of the $\eta_j$'s, and that she reports the following ``plug in'' estimate based on the $\widehat{\eta}_j$'s,
\begin{equation}\widehat{\limfunc{Var}}(\widehat{\eta}_j)=\frac{1}{p}\sum_{j=1}^p\left(\widehat{\eta}_j-\frac{1}{p}\sum_{j'=1}^p\widehat{\eta}_{j'}\right)^2.\end{equation}
Does a large estimate $\widehat{\limfunc{Var}}(\widehat{\eta}_j)$ indicate that the variance of the true effects $\eta_j$, $\limfunc{Var}(\eta_j)$, is large? Or does the presence of measurement error $v_j$ artificially inflate the dispersion in the estimates $\widehat{\eta}_j$?

As another example, suppose the researcher uses $\widehat{\eta}_j$ as a predictor of $\eta_j$. Since the $p$ parameters ${\eta}_1,...,{\eta}_p$ are estimated in the available sample and $p$ is large (or, alternatively, the noise in (\ref{eq_decomp}) is substantial), it is likely that $\widehat{\eta}_j$ ``overfits'', in the sense that it reflects too much of the noise $v_j$ and too little of the true effect $\eta_j$. In such cases, one may wish to construct predictors that lead to a lower expected sum of squared errors in (\ref{eq_MSE}).  

To illustrate these points, consider a simple setting inspired by Example \ref{Ex_1}, where $\eta_j$ and $v_j$ in (\ref{eq_decomp}) are independent, i.i.d. across $j$, and ${\cal{N}}(\mu_{\eta},\sigma_{\eta}^2)$ and ${\cal{N}}(0,\sigma_{v}^2)$, respectively. In this case, we have
\begin{equation}\mathbb{E}\left[\widehat{\limfunc{Var}}(\widehat{\eta}_j)\right]=\limfunc{Var}(\eta_j)+\frac{p-1}{p}\sigma_{v}^2,\label{eq_simple_var}\end{equation}
which shows that the variance of the estimates $\widehat{\eta}_j$ is upward-biased for the variance of the true $\eta_j$'s. Moreover, while the expected sum of squared errors of $ \widehat{\eta}_j$ is equal to $p\sigma_v^2$, the expected sum of squared errors of the conditional means \begin{equation}\mathbb{E}(\eta_j\,|\, \widehat{\eta}_j)=\frac{\sigma_{\eta}^2}{\sigma_{\eta}^2+\sigma_v^2}\widehat{\eta}_j+\frac{\sigma_{v}^2}{\sigma_{\eta}^2+\sigma_v^2}\mu_{\eta}\label{eq_simple_shrink}\end{equation} 
is smaller, equal to $\frac{p\sigma_v^2\sigma_{\eta}^2}{\sigma_{\eta}^2+\sigma_v^2}$. This shows that the ``shrunk'' quantities $\mathbb{E}(\eta_j\,|\, \widehat{\eta}_j)$ in (\ref{eq_simple_shrink}) have a lower expected sum of squared errors than the original estimates $\widehat{\eta}_j$. The difference between the two is greater when the noise variance $\sigma_v^2$ is large relative to the variance $\sigma_{\eta}^2$ of the true effects. Bias-correction methods for variance components based on equations in the spirit of (\ref{eq_simple_var}), and linear shrinkage predictors akin to (\ref{eq_simple_shrink}), are now widespread in applied economics. 

This simple example is too stylized to accurately describe the situations in Examples \ref{Ex_2} and \ref{Ex_3}, however. In such settings, estimates $\widehat{\eta}_j$ are constructed using observations from other units $j'\neq j$. Hence, the $\widehat{\eta}_j$'s are not independent. This gives rise to more complex forms for the estimation noise $v_j$ in (\ref{eq_decomp}), and complicates the way the noise affects the quantities of interest. \citet{jochmans2019fixed} study how, in settings where the matrix of covariates $z_i$ has a network structure (such as Example \ref{Ex_3}, where $z_i$ represent worker and firm employment relationships), the properties of the network, such as how connected it is, affect the precision of the estimates $\widehat{\eta}_j$. Moreover, in the settings of Examples \ref{Ex_2} and \ref{Ex_3}, unit-specific parameters $\eta_j$ may not be independent. We next present a framework that applies to an arbitrary matrix of covariates $z_i$ and allows for dependence between units.

%
%
%


%
%
%
%

\section{The normal random coefficients model\label{secModel}}

In this section we describe a normal Random Coefficients (RC) model, which allows researchers to answer the questions introduced in the previous section.

\subsection{Model and assumptions}

For the presentation we will remove the term $x_i'\beta$ from equation (\ref{eq_micro}). Depending on the setting, $\beta$ can be estimated using OLS or differenced out, see Remark \ref{rem_1} below. We then write (\ref{eq_micro}) in vector form, removing $x_i'\beta$ from the equation, as follows,
\begin{equation}
	Y=Z\eta+U,\label{eq_micro_stacked}
\end{equation}
where $Y$ is an $n\times 1 $ vector with generic element $y_i$, $Z$ is an $n\times p$ matrix with generic row $z_i'$, and $U$ is an $n\times 1 $ vector with generic element $u_i$.

The form of the design matrix $Z$ differs across applications. In Example \ref{Ex_1}, $Z$ takes the form
$$Z=\left(\begin{array}{ccccc}Z_1&0&0&...&0\\0&Z_2&0&...&0\\0&0&Z_3&...&0\\...&...&...&...&...\\0&0&0&...&Z_p\end{array}\right),$$
where $Z_j$ contains two columns, the first one being a column of $1$'s, and the second one being a column of $0$'s and $1$'s, depending on the race of the applicant. This block-diagonal structure characterizes panel data and grouped data settings.

In Examples \ref{Ex_2} and \ref{Ex_3}, $Z$ takes more complex forms. To present Example \ref{Ex_2}, let us abstract from the dependence on parental income for simplicity. Then, $z_{ij}$ in (\ref{eq_micro}) is the exposure of $i$ to neighborhood $j$. In every row of the $n\times p $ matrix with elements $z_{ij}$, all but a handful of elements are equal to zero.\footnote{For example, \citet{chetty2018impacts2} focus on families that move exactly once, so every row in the matrix has exactly two non-zero elements.} However, the matrix does not have a block-diagonal form. Then, differencing out the origin-and-destination indicators $x_i$ (as explained in Remark \ref{rem_1} below) leads to the matrix 
\begin{equation}\label{eq_Z_Ex2}Z=\left(\begin{array}{ccccc}\widetilde{z}_{11}&\widetilde{z}_{12}&\widetilde{z}_{13} & ...&\widetilde{z}_{1J}\\\widetilde{z}_{21}&\widetilde{z}_{22}&\widetilde{z}_{23} & ...&\widetilde{z}_{2J}\\\widetilde{z}_{31}&\widetilde{z}_{32}&\widetilde{z}_{33} & ...&\widetilde{z}_{3J}\\...&...&...&...&...\\\widetilde{z}_{n1}&\widetilde{z}_{n2}&\widetilde{z}_{n3} & ...&\widetilde{z}_{nJ}\end{array}\right),\end{equation}
where $\widetilde{z}_{ij}$ is equal to $z_{ij}$ minus the mean of $z_{i'j}$ for all individuals $i'$ who experience the same neighborhood moves in childhood as individual $i$ (though the time $i'$ and $i$ stay in each neighborhood may differ). $Z$ is sparse, and not block-diagonal.

In Example \ref{Ex_3}, $Z$ stacks worker and firm indicators together. For example, with two periods, $K$ workers, and $J$ firms, $Z$ reads
\begin{equation}Z=\left(\begin{array}{cccccccc}1&0&0&...&0&f_{11}^1&...&f_{11}^J\\1&0&0&...&0&f_{12}^1&...&f_{12}^J\\0&1&0&...&0&f_{21}^1&...&f_{21}^J\\0&1&0&...&0&f_{22}^1&...&f_{22}^J\\0&0&1&...&0&f_{31}^1&...&f_{31}^J\\0&0&1&...&0&f_{32}^1&...&f_{32}^J\\...&...&...&...&...&...&...&...\\0&0&0&...&1&f_{K1}^1&...&f_{K1}^J\\0&0&0&...&1&f_{K2}^1&...&f_{K2}^J\end{array}\right),\label{eq_Z_Ex3}\end{equation}
where $f_{kt}^j=1$ if worker $k$ is employed in firm $j$ in period $t$, and $f_{kt}^j=0$ otherwise. Note that $Z$ is not block-diagonal in this case. However, it is typically a sparse matrix since each row has exactly two non-zero elements. The form of $Z$ reflects the network of workers' and firms' employment relationships (\citealp{andrews2008high}, \citealp{jochmans2019fixed}).

Throughout, we assume that $Z$ has full column rank, so $(Z'Z)$ is non-singular. When $p$ is large, this assumption may be restrictive. In Example \ref{Ex_3}, ensuring non-singularity requires imposing a normalization on the parameters $\eta_j$ (e.g., that the firm-specific effects sum up to zero), and focusing on a connected component of the firm-worker network (\citealp{abowd2002computing}). Similarly, in Example \ref{Ex_2}, a normalization is needed since neighborhood effects are identified relative to the national average (\citealp{chetty2018impacts2}).

The defining assumptions for the normal random coefficients model are as follows.
\begin{assumption}{(normal RC model)}$\quad $\label{ass_nn_micro}
	
	(i) $U\,|\, Z,\eta\sim {\cal{N}}(0,\Omega(Z))$.
	
	(ii) $\eta\,|\, Z\sim {\cal{N}}(\mu(Z),\Sigma(Z))$.
\end{assumption}

In part (i) of Assumption \ref{ass_nn_micro} we assume that the error terms $u_i$ in (\ref{eq_micro}) are normally distributed, with zero mean and some $n\times n$ covariance matrix $\Omega(Z)$, independent of $\eta$. In part (ii) we specify a normal model for $\eta$ given $Z$, with mean $\mu(Z)$ (a $p\times 1$ vector) and variance $\Sigma(Z)$ (a $p\times p$ matrix). Hence, we treat the parameters $\eta_j$ as random coefficients, and we specify their conditional distribution given $Z$. This modeling device is often used in panel data and complex data settings. Note that Assumption \ref{ass_nn_micro} implies that the conditional distribution of $Y$ given $Z$ is fully specified, and normal, given the parameters $\Omega(Z)$, $\mu(Z)$, and $\Sigma(Z)$,
$$Y\,|\, Z\sim {\cal{N}}\left(Z\mu(Z),Z\Sigma(Z)Z'+\Omega(Z)\right).$$

Assumption \ref{ass_nn_micro} requires the following \emph{strict exogeneity} assumption: $\mathbb{E}[U\,|\, Z,\eta]=0$. This assumption imposes substantive restrictions on the economic environment. In Example \ref{Ex_2}, it requires that the times families spend in every neighborhoods are unrelated to the unobserved determinants of adult outcomes. In Example \ref{Ex_3}, it imposes an assumption of so-called ``exogenous mobility'', through which workers' decisions to change jobs may be driven by worker and firm effects $\eta_j$ but not by idiosyncratic time-varying shocks $u_i$. Although some authors have attempted to relax strict exogeneity in the setting of Example \ref{Ex_3} (e.g., \citealp{abowd2019modeling}, \citealp{bonhomme2019distributional}), most research to date relying on model (\ref{eq_micro}) makes this assumption.

The concrete specification of $\mu(Z)$, $\Sigma(Z)$, and $\Omega(Z)$ will depend on the application. In panel or grouped data settings, such as in Example \ref{Ex_1}, a common assumption is that $(\eta_j,Z_j)$ are independent across $j$, where $Z_j$ denotes the subset of observations $z_i$ that pertain to unit $j$. In that case, $\mu_j(Z)$ depends on $Z$ only through $Z_j$, $\Sigma(Z)$ is diagonal, and $\Sigma_{j,j}(Z)$ only depends on $Z_j$ as well. 

However, in Examples \ref{Ex_2} and \ref{Ex_3}, it may be more plausible to allow for a rich dependence of $\mu_j(Z)$ and $\Sigma_{j,j'}(Z)$ on the elements of $Z$, and to allow $\Sigma(Z)$ to be a general, non-diagonal symmetric matrix. Indeed, in Example \ref{Ex_2}, the de-meaned exposures $\widetilde{z}_{ij'}$ to other neighborhoods $j'\neq j$ in (\ref{eq_Z_Ex2}) are unlikely to be independent of neighborhood effects $\eta_j$, unless mobility across neighborhoods is unrelated to neighborhood heterogeneity. Likewise, in Example \ref{Ex_3}, indicators of firms $j'\neq j$ in (\ref{eq_Z_Ex3}) will generally correlate with the effect of firm $j$ unless workers' sorting patterns are independent of firm heterogeneity. We will return to specification issues in the next section.


Given Assumption \ref{ass_nn_micro}, the OLS estimator of $\eta$ in (\ref{eq_micro_stacked}) satisfies
\begin{equation}
	\widehat{\eta}=\eta+V,\label{eq_decomp_stacked}
\end{equation}
where, denoting the variance-covariance matrix of the OLS estimates $\widehat{\eta}$ as
\begin{equation}S(Z)=(Z'Z)^{-1}Z'\Omega(Z)Z(Z'Z)^{-1},\label{eq_S}\end{equation}
we have 
$$V=(Z'Z)^{-1}Z'U\,|\, Z,\eta\sim {\cal{N}}(0,S(Z)).$$

Model (\ref{eq_micro_stacked}) under Assumption \ref{ass_nn_micro} is a normal linear mixed model (see, e.g., \citealp{jiang2007linear} and \citealp{mcculloch2004generalized}). As we will review below, normality is not needed to obtain informative moment conditions on $\mu(Z)$, $\Sigma(Z)$, and $\Omega(Z)$, and the following restrictions on first and second moments suffice.

\begin{assumption}{(RC model)}$\quad $\label{ass_nn_micro_3}
	
	(i) $\mathbb{E}[U\,|\, Z,\eta]=0$, $\limfunc{Var}[U\,|\, Z]=\Omega(Z)$.
	
	(ii) $\mathbb{E}[\eta\,|\, Z]=\mu(Z)$, $\limfunc{Var}[\eta\,|\, Z]=\Sigma(Z)$.
	
\end{assumption}

Note that, while Assumption \ref{ass_nn_micro_3} relaxes normality, it does maintain the strict exogeneity condition $\mathbb{E}[U\,|\, Z,\eta]=0$. If the researcher is only interested in means, variances or covariances of the $\eta_j$'s, or alternatively in coefficients of regressions where $\eta_j$ appears on the left- or right--hand side, then Assumption \ref{ass_nn_micro} can be replaced by the weaker Assumption \ref{ass_nn_micro_3} that only restricts first and second moments. However, normality is needed to answer questions related to the higher-order and nonlinear moments of the $\eta_j$'s, their distributions, and to construct optimal predictors of the $\eta_j$'s.

\begin{remark}\label{rem_1}

In (\ref{eq_micro_stacked}) we have removed the term $x_i'\beta$. In applications with covariates $x_i$, there are two common ways of handling the presence of the unknown parameter $\beta$. When the dimension $q$ of $\beta$ is low, one can often reliably estimate $\widehat{\beta}$ jointly with $\widehat{\eta}$ using OLS in (\ref{eq_micro}), and then replace $Y=(y_i)$ with $Y=\left(y_i-x_i'\widehat{\beta}\right)$ in (\ref{eq_micro_stacked}). The analysis below is then essentially unchanged relative to the case without $x_i$'s. This approach is commonly used to handle the presence of age, time, and other demographics in Example \ref{Ex_3}.

In applications where $q$ is large, the parameter $\beta$ can be differenced out as follows. Write model (\ref{eq_micro}) in vector form, as
\begin{equation}
	Y=X\beta+Z\eta+U,\label{eq_micro_stacked1}
\end{equation}
where $X$ is an $n\times q $ matrix with generic row $x_i'$. Let $M$ be an $n\times n$ matrix such that $MX=0$ (e.g., a projection matrix). Left-multiplying (\ref{eq_micro_stacked1}) by $M$ then gives
\begin{equation}
	M{Y}=M{Z}\eta+M{U},\label{eq_micro_stacked2}
\end{equation}
which takes the same form as (\ref{eq_micro_stacked}). In Example \ref{Ex_2}, taking the within-group transformation as matrix $M$, this differencing approach provides a way to handle the presence of the high-dimensional origin-and-destination indicators in $x_i$.\footnote{Also, in Example \ref{Ex_3}, if one includes worker indicators in $X$, and $Z$ contain only firm indicators, then first-differencing yields an equation of the form (\ref{eq_micro_stacked2}).}

\end{remark}

\subsection{Quantities of interest in the normal RC model}

Suppose that the normal RC model (\ref{eq_micro_stacked}) holds and Assumption \ref{ass_nn_micro} is satisfied, and suppose the covariance and mean functions ${\Omega}(Z)$, ${\mu}(Z)$, and ${\Sigma}(Z)$ are known. Then the model implies closed-form expressions for all the quantities that we mentioned in Section \ref{secAnatomy}. 

For example, the first and second moments $m_c$ in (\ref{eq_lin_comb}) and $v_Q$ in (\ref{eq_quad_form}) are given, respectively, by
\begin{equation}
	m_c=\mathbb{E}\left[c'\mu(Z)\right]\label{eq_mc}
\end{equation}
and
\begin{equation}
	v_Q=\mathbb{E}\left[{\mu}(Z)'Q{\mu}(Z)+\limfunc{Trace}\left(Q{\Sigma}(Z)\right)\right].\label{eq_vQ}
\end{equation}
The expressions (\ref{eq_mc}) and (\ref{eq_vQ}) do not rely on normality, and hold under the weaker Assumption \ref{ass_nn_micro_3}.

Under Assumption \ref{ass_nn_micro}, we further can write the nonlinear moment $w_H$ in (\ref{eq_nonlin}), and the cumulative distribution function $F_{\omega}(a)$ in (\ref{eq_cdf}), in closed form as, respectively,
\begin{equation}
	{w}_H=\mathbb{E}\left[\int_{\mathbb{R}^p} H(\eta)\frac{1}{(2\pi)^{\frac{p}{2}}|{\Sigma}(Z)|^{\frac{1}{2}}}\exp\left(-\frac{1}{2}\left(\eta-{\mu}(Z)\right)'\left[{\Sigma}(Z)\right]^{-1}\left(\eta-{\mu}(Z)\right)\right)d\eta\right],
\end{equation} 
and
\begin{equation}
	F_{\omega}(a)=\mathbb{E}\left[\sum_{j=1}^p\omega_j\Phi\left(\frac{a-\mu_{j}(Z)}{\sqrt{\Sigma_{j,j}(Z)}}\right)\right],\label{eq_Fomega}
\end{equation}
where $\mu_j(Z)$ is the $j$-th element of $\mu(Z)$, $\Sigma_{j,j}(Z)$ is the $j$-th diagonal element of $\Sigma(Z)$, and $\Phi$ denotes the standard normal cumulative distribution function. Note that, while the conditional distribution of $\eta\,|\, Z$ is normal under Assumption \ref{ass_nn_micro}, the unconditional distribution of $\eta$ is not normal.

Lastly, under Assumption \ref{ass_nn_micro}, one can also derive a closed-form expression for the conditional mean of the vector $\eta$ given the data $(Y,Z)$, as
\begin{equation} \mathbb{E}\left[\eta\,|\, Y,Z\right]=G(Z)\left({S}(Z)^{-1}{\widehat{\eta}}+{\Sigma}(Z)^{-1}{\mu}(Z)\right),\label{eq_post}\end{equation}
where $S(Z)$ is given by (\ref{eq_S}), and
$$ G(Z)=\left({S}(Z)^{-1}+{\Sigma}(Z)^{-1}\right)^{-1}$$
is the conditional variance of $\eta$ given $(Y,Z)$. The conditional density of $\eta\,|\, Y,Z$ is then
\begin{equation}
	\eta\,|\, Y,Z\sim{\cal{N}}\left(\mathbb{E}\left[\eta\,|\, Y,Z\right],G(Z)\right).\label{eq_post_dens}
\end{equation}

	\section{Specification choices\label{sec_specif}}


In this section, we discuss several possibilities to specify $\Omega(Z)$, $\mu(Z)$, and $\Sigma(Z)$. 

\subsection{Specification of $\Omega(Z)$}

The simplest specification for $\Omega(Z)$ is independent homoskedastic, that is, $$\Omega(Z)=\sigma^2I_n,$$ for a constant variance parameter $\sigma^2$. In Example \ref{Ex_3}, \citet{andrews2008high} rely on this assumption and construct an unbiased estimator of $\sigma^2$ for applications to matched employer-employee data.

%
%

However, both homoskedasticity and independence can be restrictive. To relax homoskedasticity, one can introduce covariates $W=(w_i)$ (for example, some functions of the elements of $Z$), and model 
$$\Omega(Z)=\limfunc{diag}\left(\sigma_{\theta}^2(w_1),...,\sigma_{\theta}^2(w_n)\right),$$
where $\sigma_{\theta}^2(w_i)$ is a parametric function of $w_i$ indexed by some parameter $\theta$.\footnote{While here we assume that $w_i$ is a function of the elements of $Z$, the covariates $w_i$ could also contain additional covariates not functions of $Z$, such as neighborhood characteristics or worker or firm characteristics, depending on the application.} In Example \ref{Ex_2}, \citet{chetty2018impacts2} assume that $S(Z)$ in (\ref{eq_S}) is a diagonal matrix, i.e., that the estimates $\widehat{\eta}_j$ are uncorrelated across neighborhoods. Under independence, \citet{kline2020leave} model $\Omega(Z)$ as a diagonal matrix with unrestricted diagonal elements. They propose a leave-out method that provides unbiased estimates of these diagonal variance elements. 

To relax independence, one can model $\Omega(Z)$ as a parametric, matrix-valued, non-diagonal function of the covariates $w_i$. In panel data settings, \citet{arellano2012identifying} propose parametric ARMA specifications to allow for serial correlation. Relaxing independence has been shown to be important in Example \ref{Ex_3}, where assuming serial independence within employment spells is often empirically restrictive.

Lastly, it is worth noting that one cannot leave the matrix $\Omega(Z)$ fully unrestricted while at the same time identifying moments of the effects $\eta_j$. This is because $\eta_j$ and $v_j$ in (\ref{eq_decomp}) are both unobserved random variables. As a result, there is an essential trade-off between heterogeneity (the $\eta_j$'s) and the dependence of errors (the matrix $\Omega(Z)$).




\subsection{Specification of $\mu(Z)$ and $\Sigma(Z)$}

Turning now to $\mu(Z)$ and $\Sigma(Z)$, a possible approach in applications is to specify $\mu(Z)$ as a function of some covariates $W$ (e.g., as a linear function of $W$), and to model $\Sigma(Z)$ similarly (e.g., as a constant diagonal matrix). However, the cost of such an approach is that it implicitly imposes a \emph{conditional independence assumption} that may be economically restrictive. We now illustrate this important point through the help of examples.

Suppose that, in Example \ref{Ex_2}, the mean and variance of neighborhood effects are specified as functions of a handful of covariates, for instance some variables measuring the racial and economic composition of the neighborhoods. In that case, the researcher is effectively assuming that the de-meaned neighborhood exposures in $Z$, see (\ref{eq_Z_Ex2}), are independent of the location-specific effects $\eta_j$ conditional on those covariates. This assumption may be hard to reconcile with an economic model of location choice, and mobility across locations, where families' decisions are in part determined by neighborhood heterogeneity. 

Similarly, in Example \ref{Ex_3}, assuming that the means, variances, and covariances of worker and firm effects only depend on some worker and firm characteristics restricts job mobility to be independent of the worker and firm effects $\eta_j$ conditional on those characteristics. This assumption may be at odds with economic models of sorting where workers' and firms' decisions are in part determined by worker and firm heterogeneity.

To state the argument formally, let $W$ denote some covariates. Then a specification where $\mu(Z)$ and $\Sigma(Z)$ only depend on $W$ amounts to assuming the following:
$$\eta\,|\, Z,W\sim {\cal{N}}(\mu(W),\Sigma(W)),$$
which in particular imposes that:
$$Z\text{ and }\eta \text{ are independent given }W.$$
Likewise, assuming that $\mu_j(Z)=\mu_{j}(W_j)$, where $W_j$ denotes the covariates of unit $j$, imposes that:
$$\eta_j\text{ are mean of independent } \left(W_{1},...,W_{j-1},W_{j+1},...,W_p\right)\text{ given }W_j.$$


To illustrate that such conditional independence assumptions may be economically restrictive, let us consider two models of firm choice for Example \ref{Ex_3}. The first model assumes that workers maximize utility among all firms in a market, period-by-period (\citealp{card2018firms}, \citealp{lamadon2022imperfect}). Suppose that worker $k$'s indirect utility in firm $\ell$ at time $t$
is
$$V_{k\ell t}=\rho W_{k \ell t}+\varepsilon_{k\ell t},$$
where $\varepsilon_{k\ell t}$ are i.i.d. type I extreme value preference shocks, independent of log wages $W_{k \ell t}$, and we abstract from non-wage amenities for simplicity. Then the probability that worker $k$ chooses firm $\ell$, given all wages $W=\{W_{k'\ell' t}\}$, is
\begin{equation}\Pr\left(\ell\,|\, k, W\right)=\frac{\exp\left(\rho W_{k \ell t}\right)}{\sum_{\ell'\in {\cal{M}}(k)}\exp\left(\rho W_{k \ell' t}\right)},\label{eq_mob_static}\end{equation}
where ${\cal{M}}(k)$ is the market that $k$ considers when looking for a job. Therefore, the elements in the $Z$ matrix depend on the log wages $W_{k \ell t}$. If, further, log wages are a function of worker heterogeneity $\alpha_k$ and firm heterogeneity $\psi_{\ell}$,\footnote{For example, if log wages are given by the additive specification $W_{k \ell t}=\alpha_k+\psi_{\ell}+U_{k\ell t}$ of \citet{abowd1999high}, in which case $\Pr\left(\ell\,|\, k, W\right)$ in (\ref{eq_mob_static}) does not depend on $k$.}
then (\ref{eq_mob_static}) implies that the $\eta_j$'s, which here are the $\alpha_k$'s and $\psi_{\ell}$'s, are \emph{not} independent of $Z$, even conditional on observed characteristics.     

Consider next a dynamic model of workers' mobility across firms, as proposed by \citet{lentz2023anatomy} (see \citealp{sorkin2018ranking} for a related model). The probability that worker $k$ moves between firms $\ell$ and $\ell'$, conditional on worker heterogeneity $\alpha=\{\alpha_k\}$ and firm heterogeneity $\psi=\{\psi_{\ell}\}$, is specified as
\begin{equation}\Pr\left(\ell'\,|\, \ell,k,\alpha,\psi\right)=\lambda_{k\ell'}\frac{\gamma_{k\ell'}}{\gamma_{k\ell}+\gamma_{k\ell'}},\label{eq_mob_dyn}\end{equation}
where $\lambda_{k\ell'}$ is the probability that $k$ meets firm $\ell'$, and $\gamma_{k\ell}$ is interpreted as worker $k$'s value of working in firm $\ell$. If the value $\gamma_{k\ell}=\gamma(\alpha_k,\psi_{\ell})$ depends on worker heterogeneity $\alpha_k$ and firm heterogeneity $\psi_{\ell}$, then neither $\alpha_k$ nor $\psi_{\ell}$ are independent of $Z$, even conditional on observed characteristics.



We now discuss several examples of specifications for $\mu(Z) $ and $\Sigma(Z)$ used in practice. In Example \ref{Ex_2}, a common approach is to model $\mu_j(Z)$ to be a linear function of some covariates $W_j$, and $\Sigma(Z)$ to be a diagonal matrix, independent of $Z$ and $W$. The model then assumes independence across $j$'s, and rules out dependence between the true effects ($\eta_j$) and location choice and mobility ($Z$) conditional on the covariates. Recently, \citet{chen2023empirical} proposes an extension of this approach that allows for dependence between the true effects $\eta_j$ and the precision of their estimates $\widehat{\eta}_j$ (as measured by the diagonal elements of $S(Z)$ in (\ref{eq_S})), while maintaining independence across units.

In Example \ref{Ex_3}, \citet{woodcock2015match} postulates a normal RC model where neither $\mu(Z)$ nor $\Sigma(Z)$ depend on $Z$, and $\Sigma(Z)$ is a diagonal matrix. However, these assumptions impose that workers' sorting patterns, which are encoded in $Z$, do not depend on the worker and firm effects $\eta_j$. \citet{bonhomme2023much} refine the \citet{woodcock2015match} model by allowing $\mu(Z)$ and $\Sigma(Z)$ to depend on $Z$. To model the dependence, they cluster firms into a small number of groups using the k-means algorithm based on their wage distributions (as in \citealp{bonhomme2019distributional}). Given this grouping, they allow the means and variances of worker and firm effects to
depend on the groups, but not on the worker and firm identities within these groups. Similarly, they allow the covariances in $\Sigma(Z)$, which they do not assume to be diagonal, to depend on the groups. Generalizing such approaches to accommodate structural economic models of workers' mobility across firms is a promising area for future investigation.

\section{Estimation\label{secIdent}}

In this section we describe various estimation strategies for the parameters of the normal RC model and the quantities of interest introduced in Section \ref{secAnatomy}.

\subsection{Estimating the model's parameters}

We start by providing moment conditions on the primitive parameters of the normal RC model: the error variance $\Omega(Z)$, and the mean and variance of the unit-specific effects $\mu(Z)$ and $\Sigma(Z)$.

\paragraph{Moment conditions.}

Let Assumption \ref{ass_nn_micro_3} hold. Then we have\begin{equation}
	\mathbb{E}\left[YY'\,|\, Z\right]=Z\mathbb{E}\left[\eta\eta'\,|\, Z\right]Z'+\Omega(Z),\label{eq_mat_omega}\end{equation}
where the cross-product term is zero since $\mathbb{E}[U\,|\, Z,\eta]=0$. This is a system of $n\times n$ moment conditions. Following \citet{arellano2012identifying}, we can construct a suitable $n^2\times n^2$ matrix $M(Z)$ that ``differences out'' the first term on the right-hand side of (\ref{eq_mat_omega}),\footnote{Let $A\otimes B$ denote the Kronecker product between $A$ and $B$, and let $I_{n^2}$ be the $n^2\times n^2$ identity matrix. A possible choice is $M(Z)=I_{n^2}-\left(Z(Z'Z)^{-1}Z'\right)\otimes \left(Z(Z'Z)^{-1}Z'\right)$. The key property is that $M(Z)(Z\otimes Z)=0$, which implies that $M(Z)\limfunc{vec}\left(Z\mathbb{E}\left[\eta\eta'\,|\, Z\right]Z'\right)=M(Z)(Z\otimes Z)\limfunc{vec}\left(\mathbb{E}\left[\eta\eta'\,|\, Z\right]\right)=0$.} to obtain 
\begin{equation}\mathbb{E}\left[M(Z)\limfunc{vec}\left(YY'-\Omega(Z)\right)\,|\, Z\right]=0.\label{eq_vec_omega3}\end{equation}
where the vec operator stacks together the $n$ columns of an $n\times n$ matrix into a $n^2\times 1$ vector. (\ref{eq_vec_omega3}) provides a system of conditional moment conditions on the elements of $\Omega(Z)$. Note that, depending on the specification of $\Omega(Z)$, the moment conditions (\ref{eq_vec_omega3}) do not necessarily guarantee that the elements of $\Omega(Z)$ are identified. Note also that $M(Z)$ in (\ref{eq_vec_omega3}) being singular implies that the system of equations has multiple solutions in the absence of restrictions on $\Omega(Z)$.

Next, taking the conditional mean in (\ref{eq_decomp_stacked}), we obtain
\begin{equation}\mathbb{E}[\widehat{\eta}-\mu(Z)\,|\, Z]=0,\label{eq_mean_restr}\end{equation}
which provide conditional moment conditions on $\mu(Z)$. These moment conditions depend neither on $\Omega(Z)$ nor on $\Sigma(Z)$.

Lastly, using (\ref{eq_decomp_stacked}) we obtain
\begin{align*}\limfunc{Var}[\widehat{\eta}\,|\, Z]=\Sigma(Z)+S(Z),\end{align*}
where $S(Z)$ given by (\ref{eq_S}) is a function of $\Omega(Z)$. This gives the following moment conditions on $\Sigma(Z)$, $\Omega(Z)$, and $\mu(Z)$,
\begin{equation}\mathbb{E}[\left(\widehat{\eta}-\mu(Z)\right)\left(\widehat{\eta}-\mu(Z)\right)'-S(Z)-\Sigma(Z)\,|\, Z]=0.\label{eq_var_restr2}\end{equation}

\begin{remark}
In general, (\ref{eq_vec_omega3})-(\ref{eq_mean_restr})-(\ref{eq_var_restr2}) do not exhaust all the information about $\Omega(Z)$, $\mu(Z)$, and $\Sigma(Z)$. The complete set of first- and second-moment conditions implied by Assumption \ref{ass_nn_micro_3} consists of (\ref{eq_mean_restr}) and
\begin{align}
	&\mathbb{E}\left[YY'\,|\, Z\right]=Z\left(\Sigma(Z)+\mu(Z)\mu(Z)'\right)Z'+\Omega(Z).\label{eq_complete}
\end{align}
An advantage of the moment conditions in (\ref{eq_vec_omega3}) is that they are robust to possible misspecification of $\mu(Z)$ and $\Sigma(Z)$.
\end{remark}

\paragraph{Parameter estimation.}

Given a parametric or semi-parametric specification for $\Omega(Z)$, $\mu(Z)$, and $\Sigma(Z)$, a possible estimation approach is to exploit the moment conditions (\ref{eq_vec_omega3})-(\ref{eq_mean_restr})-(\ref{eq_var_restr2}) using method-of-moments or minimum-distance estimation. This strategy is used in \cite{bonhomme2023much} in Example \ref{Ex_3}, for instance. Let $\theta$ be a parameter vector indexing $\mu_{\theta}(Z)$, $\Sigma_{\theta}(Z)$, and $\Omega_{\theta}(Z)$, possibly adding other conditioning covariates $W$. This estimation step delivers an estimate $\widehat{\theta}$, as well as estimates $\widehat{\mu}(Z)=\mu_{\widehat{\theta}}(Z)$, $\widehat{\Sigma}(Z)=\Sigma_{\widehat{\theta}}(Z)$, and $\widehat{\Omega}(Z)=\Omega_{\widehat{\theta}}(Z)$. An alternative is to perform (quasi-) maximum likelihood estimation, as in the following remark. 

	\begin{remark}\label{rem_3}

	An alternative approach to estimation of $\mu$, $\Sigma$ and $\Omega$ is to rely on the log-likelihood function conditional on $Z$, 
	\begin{align}L(\mu,\Sigma,\Omega)&=-\frac{n}{2}\log\left(2\pi\right)-\frac{1}{2}\log\left(|Z\Sigma(Z)Z'+\Omega(Z)|\right)\notag\\
		&\quad \quad \quad -\frac{1}{2}\left(Y-Z\mu(Z)\right)'\left[Z\Sigma(Z)Z'+\Omega(Z)\right]^{-1}\left(Y-Z\mu(Z)\right).\label{eq_L2}
	\end{align}
	In the absence of normality, i.e., under Assumption \ref{ass_nn_micro_3}, $L(\mu,\Sigma,\Omega)$ can be interpreted as a quasi log-likelihood function.\footnote{Moreover, when based on $A'Y$ and $A'Z$, where $M=AA'$ is the Cholesky decomposition of the matrix $M$ of Remark \ref{rem_1}, (\ref{eq_L2}) is the basis for restricted maximum likelihood (REML) estimation, and the resulting REML estimator does not depend on the choice of $A$ (see, e.g., \citealp{jiang2007linear}).}

	%
\end{remark}

%

%




%
%

%
%

%

	\subsection{Estimating the quantities of interest: three strategies\label{subsec_PAE}}

	We now present three types of estimators for the quantities of interest introduced in Section \ref{secAnatomy}.

	\paragraph{\#1 Bias-corrected fixed-effects estimators.}

	Suppose first that the researcher is only interested in estimating linear combinations of the $\eta_j$'s or quadratic forms. In that case, she does not need to estimate $\mu(Z)$ and $\Sigma(Z)$. Indeed, linear combinations $m_c=\mathbb{E}\left[c'\eta\right]$ and, given knowledge of $\Omega(Z)$, quadratic forms $v_Q=\mathbb{E}[{\eta}'Q\eta]$, are nonparametrically identified under Assumption \ref{ass_nn_micro_3} (without the need for normality), as
	\begin{align}
		m_c=\mathbb{E}[c'\widehat{\eta}]\label{eq_mean_BC}
	\end{align}
	and
	\begin{align}
		v_Q=\mathbb{E}[\widehat{\eta}'Q\widehat{\eta}]-\mathbb{E}\left[\limfunc{Trace}\left(QS(Z)\right)\right],\label{eq_quad_form_BC}
	\end{align}
	respectively. By (\ref{eq_mean_BC}), linear combinations of the elements in $\widehat{\eta}$ are unbiased. Moreover, (\ref{eq_quad_form_BC}) shows that, while quadratic forms in $\widehat{\eta}$ are biased, the bias is a known function of the variance-covariance matrix $S(Z)$ of $\widehat{\eta}$.

	%
	%


	Then, a fixed-effects estimator of $m_c$ is
	\begin{equation}
		\widehat{m}^{\rm FE}_c=c'\widehat{\eta}.\label{eq_mean_BC_FE}
	\end{equation}
Moreover, given an estimator $\widehat{\Omega}(Z)$ and an associated estimator $\widehat{S}(Z)$ given by 
	\begin{equation}\widehat{S}(Z)=(Z'Z)^{-1}Z'\widehat{\Omega}(Z)Z(Z'Z)^{-1},\label{eq_Shat}\end{equation} a bias-corrected fixed-effects estimator of $v_Q$ is
	\begin{align}
		\widehat{v}^{\rm FE}_Q=\widehat{\eta}'Q\widehat{\eta}-\limfunc{Trace}\left(Q\widehat{S}(Z)\right).\label{eq_quad_form_BC_FE}
	\end{align}
	Under Assumption \ref{ass_nn_micro_3}, the estimator $\widehat{v}^{\rm FE}_Q$ is unbiased whenever $\widehat{\Omega}(Z)$, and hence $\widehat{S}(Z)$, are themselves unbiased.

In Example \ref{Ex_3}, \citet{andrews2008high} assume that $\Omega(Z)=\sigma^2I_n$. In this case, (\ref{eq_vec_omega3}) is equivalent to
$$\mathbb{E}\left[\left(I_n-Z(Z'Z)^{-1}Z'\right)YY'\left(I_n-Z(Z'Z)^{-1}Z'\right)\,|\, Z\right]=\sigma^2\left(I_n-Z(Z'Z)^{-1}Z'\right).$$
So, by taking the trace and expectation with respect to $Z$, it follows that
\begin{equation}\sigma^2=\mathbb{E}\left[\frac{Y'\left(I_n-Z(Z'Z)^{-1}Z'\right)Y}{n-p}\right].\label{eq_sigma}\end{equation}
The formula (\ref{eq_sigma}) is the well-known degree of freedom correction for variance estimation. In Example \ref{Ex_3}, \citet{andrews2008high} propose the estimators
\begin{equation}\widehat{\sigma}^2=\frac{Y'\left(I_n-Z(Z'Z)^{-1}Z'\right)Y}{n-p}\label{eq_sigma_hat}
\end{equation}
and
\begin{align}
	\widehat{v}^{\rm FE}_Q=\widehat{\eta}'Q\widehat{\eta}-\widehat{\sigma}^2\limfunc{Trace}\left(Q(Z'Z)^{-1}\right),\label{eq_quad_form_BC_FE2}
\end{align}
which are unbiased for $\sigma^2$ and $v_Q$, respectively. \citet{kline2020leave} generalize their approach to the case where $\Omega(Z)$ is a diagonal matrix with unrestricted diagonal elements.

	\paragraph{\#2 Model-based estimators.}
	
Suppose now that the researcher wishes to estimate not only linear combinations and quadratic forms but also other quantities, such as distributions, nonlinear moments, or predictors. In that case, she first needs to produce estimates $\widehat{\Omega}(Z)$, $\widehat{\mu}(Z)$, and $\widehat{\Sigma}(Z)$. Given those, the researcher can produce estimates of all the quantities of interest we listed in Section \ref{secAnatomy}, as follows:
	\begin{align}
		&\widehat{m}_c=c'\widehat{\mu}(Z),\label{eq_mc_est}\\
		&\widehat{v}_Q=\widehat{\mu}(Z)'Q\widehat{\mu}(Z)+\limfunc{Trace}\left(Q\widehat{\Sigma}(Z)\right),\label{eq_Q_est}\\
		&\widehat{w}_H=\int_{\mathbb{R}^p}H(\eta)\frac{1}{(2\pi)^{\frac{p}{2}}|\widehat{\Sigma}(Z)|^{\frac{1}{2}}}\exp\left(-\frac{1}{2}\left(\eta-\widehat{\mu}(Z)\right)'\left[\widehat{\Sigma}(Z)\right]^{-1}\left(\eta-\widehat{\mu}(Z)\right)\right)d\eta,\label{eq_wH_est}\\
	&\widehat{F}_{\omega}(a)=\sum_{j=1}^p\omega_j\Phi\left(\frac{a-\widehat{\mu}_{j}(Z)}{\sqrt{\widehat{\Sigma}_{j,j}(Z)}}\right),\label{eq_F_est}
\end{align}
including estimates of the posterior mean and variance of $\eta$:
	\begin{align}
	&	\widehat{\mathbb{E}}\left[\eta\,|\,Y,Z\right]=\widehat{G}(Z)\left(\widehat{S}(Z)^{-1}{\widehat{\eta}}+\widehat{\Sigma}(Z)^{-1}\widehat{\mu}(Z)\right),\label{eq_PM_est}\\
	&	 \widehat{G}(Z)=\left(\widehat{S}(Z)^{-1}+\widehat{\Sigma}(Z)^{-1}\right)^{-1},\label{eq_postvar_est}
	\end{align}
	where $\widehat{S}(Z)$ is given by (\ref{eq_Shat}).

\paragraph{\#3 Posterior estimators.}

To motivate the third type of estimators, consider a generic nonlinear moment of $\eta$, ${w}_H=\mathbb{E}\left[H(\eta)\right]$, and note that, by the law of iterated expectations,
\begin{align}
	{w}_H&=\mathbb{E}\left[\mathbb{E}\left(H(\eta)\,|\, Y,Z\right)\right]\notag\\
	&=\mathbb{E}\left[\int_{\mathbb{R}^p} H(\eta)\frac{1}{(2\pi)^{\frac{p}{2}}|{G}(Z)|^{\frac{1}{2}}}\exp\left(-\frac{1}{2}\left(\eta-\mathbb{E}[\eta\,|\, Y,Z]\right)'\left[G(Z)\right]^{-1}\left(\eta-\mathbb{E}[\eta\,|\, Y,Z]\right)\right)d\eta\right],\label{eq_inter}
\end{align}
where we have used the expression in (\ref{eq_post_dens}) of the conditional density of $\eta\,|\, Y,Z$.

Equation (\ref{eq_inter}) motivates the following posterior estimator of $w_H$:
\begin{equation}
	\widehat{w}^{\rm POST}_H=\int_{\mathbb{R}^p} H(\eta)\frac{1}{(2\pi)^{\frac{p}{2}}|\widehat{G}(Z)|^{\frac{1}{2}}}\exp\left(-\frac{1}{2}\left(\eta-\widehat{\mathbb{E}}[\eta\,|\, Y,Z]\right)'\left[\widehat{G}(Z)\right]^{-1}\left(\eta-\widehat{\mathbb{E}}[\eta\,|\, Y,Z]\right)\right)d\eta,
\end{equation}
where $\widehat{\mathbb{E}}[\eta\,|\, Y,Z]$ and $\widehat{G}(Z)$ are given by (\ref{eq_PM_est}) and (\ref{eq_postvar_est}), respectively.

To provide intuition about the posterior estimator, we note that, by a change in variables,
\begin{equation}
	\widehat{w}^{\rm POST}_H= \mathbb{E}\left[H\left(\widehat{\mathbb{E}}[\eta\,|\, Y,Z]+\widehat{G}(Z)^{\frac{1}{2}}\varepsilon\right)\,|\, Y,Z\right],\label{eq_CIV}
\end{equation}
where $\varepsilon\,|\, Y,Z\sim {\cal{N}}(0,I_p)$. Consider two special cases in (\ref{eq_CIV}). When the $\widehat{\eta}_j$ are poorly estimated so $\widehat{S}(Z)^{-1}\approx 0$, then (\ref{eq_CIV}) becomes 
\begin{equation}
	\widehat{w}^{\rm POST}_H\approx \mathbb{E}\left[H\left(\widehat{\mu}(Z)+\widehat{\Sigma}(Z)^{\frac{1}{2}}\varepsilon\right)\,|\, Z\right],\label{eq_CIV_3}
\end{equation}
which coincides with the model-based estimator (\ref{eq_wH_est}). Now, when the $\widehat{\eta}_j$ are well estimated so $\widehat{S}(Z)\approx 0$ (and $S(Z)\approx 0$), then (\ref{eq_CIV}) becomes
\begin{equation}
	\widehat{w}^{\rm POST}_H\approx H\left(\widehat{\eta}\right)\approx H\left(\eta\right),\label{eq_CIV_2}
\end{equation}
which means that the posterior estimator recovers the true quantity in this situation, irrespective of whether the normal RC model is well specified or not. This illustrates a robustness property of posterior estimators, as studied by \citet{arellano2009robust} and \citet{bonhomme2022posterior}, and provides a motivation for reporting such estimators in practice.

\subsection{Remarks on asymptotic properties and inference}

	There are several challenges to deriving asymptotic properties for the estimators mentioned in this section, in the type of settings we are focusing on in this paper. A first challenge is the dimensionality of the data and model. Typically, the vector $\mu(Z)$ and the matrices $\Omega(Z)$ and $\Sigma(Z)$ are very large (i.e., both $n$ and $p$ are large). It is therefore necessary to work in a high-dimensional asymptotic regime where both $n$ and $p$ tend to infinity.  Moreover, the specifications for $\Omega(Z)$, $\mu(Z)$, and $\Sigma(Z)$ often depend on many parameters. Another challenge is the nature of the matrix $Z$. In Examples \ref{Ex_2} and \ref{Ex_3}, $Z$ does not have a block-diagonal structure, which further complicates the asymptotic analysis. Related to this, $\Sigma(Z)$ is often non-diagonal, hence creating complex forms of dependence among observations.

To give a simple example, suppose we are interested in the mean $m_c=\mathbb{E}[\frac{1}{p}\sum_{j=1}^p\eta_j]$. Consistency of $\widehat{m}_c=\frac{1}{p}\sum_{j=1}^p\widehat{\mu}_j(Z)$ for $m_{c}$ is not immediate. To see this, consider first the case where there is no estimation error and one can compute $\widetilde{m}_c=\frac{1}{p}\sum_{j=1}^p\mu_j(Z)$. Consistency of $\widetilde{m}_c$ for $m_{c}$ requires limiting the dependence of the $\mu_{j}(Z)$'s across $j$. For instance, in Example \ref{Ex_2}, this requires limiting the dependence of mean exposure effects across neighborhoods, while in Example \ref{Ex_3}, this requires limiting the dependence of mean firm effects across different firms, for example, allowing for dependence among ``similar'' firms only. Next, the addition of estimation error further complicates the argument, since the $\widehat{\mu}_j(Z)$'s are dependent across $j$ conditional on $Z$.

Nevertheless, several results are available in the literature. For the case of fixed-effects estimators of quadratic forms, as in (\ref{eq_quad_form_BC_FE}), \citet{kline2020leave} provide conditions for consistency, as well as inference theory, in a setup that assumes $\Omega(Z)$ to be diagonal while leaving $\mu(Z)$ and $\Sigma(Z)$ unrestricted. 

The mixed models literature in statistics (as reviewed by, e.g., \citealp{jiang2017asymptotic}) provides a variety of asymptotic results for the quantities of interest we have listed here, including for estimators of posterior means such as (\ref{eq_PM_est}). However, the models studied in the theoretical literature on mixed models impose specific assumptions on $\Omega(Z)$, $\mu(Z)$, and $\Sigma(Z)$. In particular, a typical assumption is that $\Sigma(Z)$ is a diagonal matrix, as in the so-called ``mixed ANOVA'' model. We have argued here that such an assumption may be economically restrictive. For example, in Example \ref{Ex_3}, $\Sigma(Z)$ is not diagonal whenever the effects of two firms $j$ and $j'$ depend on each other given $Z$, as happens in the model of \citet{lentz2023anatomy} mentioned in Section \ref{sec_specif}. 

Under Assumption \ref{ass_nn_micro}, which imposes normality, \citet{heijmans1986consistent} provide conditions for consistency of the maximum likelihood estimator (as in Remark \ref{rem_3}) in models with general forms of dependence across observations induced by the matrices $\Omega(Z)$ and $\Sigma(Z)$. \citet{heijmans1986asymptotic} provide a further set of conditions under which the maximum likelihood estimator is asymptotically normal. Importantly, however, while their setup allows for general forms of dependence, it relies on the assumption that the dimension of the model's parameter $\theta$ is kept fixed as the sample size tends to infinity.

To make the model more flexible while keeping the number of parameters to estimate moderate, a possibility is to assume that, while means and variances vary across observations, this variation is driven by a small number of groups. For example, \citet{bonhomme2023much} assume that the means and variances of worker and firm effects depend on firm groups. Grouping methods, where group membership is estimated using methods akin to the k-means algorithm, are theoretically justified under suitable conditions (\citealp{bonhomme2015grouped}, \citealp{bonhomme2022discretizing}). Assuming a grouped structure reduces the dimension of the model, and can allow for a simpler characterization of asymptotic distributions.    

Given the relevance of these high-dimensional, dependent settings for the applied economic literature, more research needs to be done regarding the formal analysis of asymptotic properties and inference methods, both in cases where normality is assumed to hold (as in Assumption \ref{ass_nn_micro}) and in cases where distributions may be non-normal. We next comment on the possibility that normality may be violated.

	\section{Extensions\label{sec_relax}}

While the normal random coefficients model provides a simple and powerful tool to estimate heterogeneous effects in high-dimensional settings, it relies on functional forms assumptions that may be empirically restrictive. In this final section of the paper, we briefly explore some strategies that could be used to relax some of those assumptions.

\paragraph{Non-normal effects.}

  When it is suspected that the distribution of $\eta\,|\, Z$ is not truly normal, that distribution is sometimes interpreted as a Bayesian prior. This interpretation is central to the empirical Bayes approach, and it is often invoked in applications such as Example \ref{Ex_2}. In this perspective, the conditional distribution of $\eta\,|\, Y,Z$ in (\ref{eq_post_dens}) can be interpreted as the posterior distribution of $\eta$. Linear shrinkage predictors as in (\ref{eq_PM_est}) possess attractive robustness properties under misspecification (\citealp{james1961estimation}).\footnote{In particular, the James-Stein linear shrinkage estimator in the normal means model achieves asymptotically minimax expected sum-of-squares loss in Euclidean balls (see Theorem 7.48 in \citealp{wasserman2006all}).} Moreover, estimators of nonlinear moments based on posterior distributions (see Subsection \ref{subsec_PAE}) are less sensitive to misspecification of the normal model compared to estimators based on the prior distribution $\eta\,|\, Z$.  However, in many applications the variance of the noise is substantial, and it is still important to correctly model the distribution of $\eta\,|\, Z$.
  


Suppose that we maintain the normality of $U\,|\, Z$ in Assumption \ref{ass_nn_micro}, while leaving the conditional distribution of $\eta\,|\, Z$ unrestricted. Then, (\ref{eq_decomp}) becomes a nonparametric deconvolution model with normal errors. Identification and estimation strategies exist in a variety of settings. In Example \ref{Ex_1}, \citet{kline2022systemic} use Efron's deconvolution method (\citealp{efron2016empirical}) to estimate the density of firm-specific discrimination under an independence assumption. \citet{chen2023empirical} proposes a location-scale model to handle situations where the precision of the estimates $\widehat{\eta}_j$ predicts the true $\eta_j$, assuming independence across units. He shows that this modeling can improve the prediction performance of conditional mean estimates. Relaxing independence across units, thereby fitting the settings of Examples \ref{Ex_2} and \ref{Ex_3}, is an important task for future work on these approaches.

\paragraph{Non-normal noise.}

 One may also suspect that the distribution of $U\,|\, Z$ is not normal. In applications to Examples \ref{Ex_1} and \ref{Ex_2}, the error $V$ in (\ref{eq_decomp_stacked}) may be approximately normal even when $U\,|\, Z$ is not, due to the fact that $V$ is an estimation error, hence asymptotically normal under standard conditions as the relevant sample size tends to infinity. Approximate normality of $V$ will typically hold in grouped data settings as Example \ref{Ex_1} provided that group sizes be large enough.\footnote{For example, $V$  may be approximately normal even when $Y$ is a vector of binary outcomes. In \citet{kline2022systemic}, the outcomes $y_i$ are binary call-back indicators, yet estimates $\widehat{\eta}_j$ may still be approximately normal with variance $S(Z)$.} However, the normal approximation need not be accurate in other applications.\footnote{When $U\,|\, Z$ is not normal, the right-hand side in (\ref{eq_post}) remains unbiased whenever $U$ has zero mean given $Z$ and $\eta$. However, it is no longer the best predictor of $\eta$ in general.}

Dealing with settings where $V$ in (\ref{eq_decomp_stacked}) is not normal, and $\eta\,|\, Z$ is not normal either, is challenging. This requires using the individual outcome model (\ref{eq_micro}), while allowing for a non-normal distribution of $U\,|\ Z$. In panel data settings, \citet{arellano2012identifying} show how to identify and estimate the distribution of $U$ under the assumption that it follows a linear independent factor model (see also \citealp{kotlarski1967characterizing}, \citealp{li1998nonparametric}, and \citealp{bonhomme2010generalized}). We are not aware of extensions of such generalized deconvolution approaches to the settings of Examples \ref{Ex_2} and \ref{Ex_3}, however.   

\paragraph{Nonlinear mean.}

Lastly, an important assumption in the RC model is that the mean outcome is linear in $\eta$, in the sense that 
\begin{equation}
	\mathbb{E}\left[Y\,|\, Z,\eta\right]=Z\eta.\label{eq_linearmean}
\end{equation}
Equation (\ref{eq_linearmean}) has empirical and economic content. In Example \ref{Ex_2}, assuming that neighborhood exposures experienced by children have a separable, constant impact on outcomes may be restrictive (\citealp{chetty2018impacts2}). In Example \ref{Ex_3}, assuming away interactions between worker and firm effects in wages implicitly imposes restrictive assumptions on input complementarity, which in turn drives the nature of sorting patterns in many economic models (e.g., \citealp{becker1973theory}, \citealp{eeckhout2011identifying}).

Within the context of Example \ref{Ex_3}, \citet{bonhomme2019distributional} show how to allow for complementarity patterns between firm and worker effects, in a setup where they assume that firm heterogeneity is discrete.\footnote{In that application, the presence of complementarity between worker and firm effects can alternatively be interpreted as reflecting individual heterogeneity in the ``treatment effect'' of a firm, as studied in \citet{hull2018estimating}.} More work is needed in this direction.

	\clearpage

{\small
	\bibliographystyle{econometrica}
	\bibliography{biblio}

@article{kline2022systemic,
	title={Systemic discrimination among large US employers},
	author={Kline, Patrick and Rose, Evan K and Walters, Christopher R},
	journal={The Quarterly Journal of Economics},
	volume={137},
	number={4},
	pages={1963--2036},
	year={2022},
	publisher={Oxford University Press}
}

@article{chetty2018impacts,
	title={The impacts of neighborhoods on intergenerational mobility I: Childhood exposure effects},
	author={Chetty, Raj and Hendren, Nathaniel},
	journal={The Quarterly Journal of Economics},
	volume={133},
	number={3},
	pages={1107--1162},
	year={2018},
	publisher={Oxford University Press}
}

@article{chetty2018impacts2,
	title={The impacts of neighborhoods on intergenerational mobility II: County-level estimates},
	author={Chetty, Raj and Hendren, Nathaniel},
	journal={The Quarterly Journal of Economics},
	volume={133},
	number={3},
	pages={1163--1228},
	year={2018},
	publisher={Oxford University Press}
}

@article{abowd1999high,
	title={High wage workers and high wage firms},
	author={Abowd, John M and Kramarz, Francis and Margolis, David N},
	journal={Econometrica},
	volume={67},
	number={2},
	pages={251--333},
	year={1999},
	publisher={Wiley Online Library}
}

@techreport{bergman2019creating,
	title={Creating moves to opportunity: Experimental evidence on barriers to neighborhood choice},
	author={Bergman, Peter and Chetty, Raj and DeLuca, Stefanie and Hendren, Nathaniel and Katz, Lawrence F and Palmer, Christopher},
	year={2019},
	institution={National Bureau of Economic Research}
}

@article{chamberlain1992efficiency,
	title={Efficiency bounds for semiparametric regression},
	author={Chamberlain, Gary},
	journal={Econometrica: Journal of the Econometric Society},
	pages={567--596},
	year={1992},
	publisher={JSTOR}
}

@article{arellano2012identifying,
	title={Identifying distributional characteristics in random coefficients panel data models},
	author={Arellano, Manuel and Bonhomme, St{\'e}phane},
	journal={The Review of Economic Studies},
	volume={79},
	number={3},
	pages={987--1020},
	year={2012},
	publisher={Oxford University Press}
}

@article{chamberlain2009decision,
	title={Decision theory applied to a linear panel data model},
	author={Chamberlain, Gary and Moreira, Marcelo J},
	journal={Econometrica},
	volume={77},
	number={1},
	pages={107--133},
	year={2009},
	publisher={Wiley Online Library}
}

@article{chamberlain2016fixed,
	title={Fixed Effects, Invariance, and Spatial Variation in Intergenerational Mobility},
	author={Chamberlain, Gary},
	journal={American Economic Review},
	volume={106},
	number={5},
	pages={400--404},
	year={2016},
	publisher={American Economic Association 2014 Broadway, Suite 305, Nashville, TN 37203}
}

@book{jiang2007linear,
	title={Linear and generalized linear mixed models and their applications},
	author={Jiang, Jiming and Nguyen, Thuan},
	volume={1},
	year={2007},
	publisher={Springer}
}

@book{mcculloch2004generalized,
	title={Generalized, linear, and mixed models},
	author={McCulloch, Charles E and Searle, Shayle R},
	year={2004},
	publisher={John Wiley \& Sons}
}

@article{aloni2023one,
	title={One Land, Many Promises: The Unequal Consequences of Childhood Location for Natives and Immigrants in Israel},
	author={Aloni, Tslil and Avivi, Hadar},
	year={2023}
}

@article{andrews2008high,
	title={High wage workers and low wage firms: negative assortative matching or limited mobility bias?},
	author={Andrews, Martyn J and Gill, Len and Schank, Thorsten and Upward, Richard},
	journal={Journal of the Royal Statistical Society Series A: Statistics in Society},
	volume={171},
	number={3},
	pages={673--697},
	year={2008},
	publisher={Oxford University Press}
}

@article{kline2020leave,
	title={Leave-out estimation of variance components},
	author={Kline, Patrick and Saggio, Raffaele and S{\o}lvsten, Mikkel},
	journal={Econometrica},
	volume={88},
	number={5},
	pages={1859--1898},
	year={2020},
	publisher={Wiley Online Library}
}

@article{robinson1991blup,
	title={That BLUP is a good thing: the estimation of random effects},
	author={Robinson, George K},
	journal={Statistical science},
	pages={15--32},
	year={1991},
	publisher={JSTOR}
}

@techreport{abowd2002computing,
	title={Computing person and firm effects using linked longitudinal employer-employee data},
	author={Abowd, John M and Creecy, Robert H and Kramarz, Francis},
	year={2002},
	institution={Center for Economic Studies, US Census Bureau}
}

@article{bonhomme2023much,
	title={How much should we trust estimates of firm effects and worker sorting?},
	author={Bonhomme, St{\'e}phane and Holzheu, Kerstin and Lamadon, Thibaut and Manresa, Elena and Mogstad, Magne and Setzler, Bradley},
	journal={Journal of Labor Economics},
	volume={41},
	number={2},
	pages={291--322},
	year={2023},
	publisher={The University of Chicago Press Chicago, IL}
}

@article{chen2023empirical,
	title={Empirical Bayes When Estimation Precision Predicts Parameters},
	author={Chen, Jiafeng},
	year={2023}
}

@article{woodcock2015match,
	title={Match effects},
	author={Woodcock, Simon D},
	journal={Research in Economics},
	volume={69},
	number={1},
	pages={100--121},
	year={2015},
	publisher={Elsevier}
}

@incollection{bonhomme2020econometric,
	title={Econometric analysis of bipartite networks},
	author={Bonhomme, St{\'e}phane},
	booktitle={The econometric analysis of network data},
	pages={83--121},
	year={2020},
	publisher={Elsevier}
}

@article{bonhomme2019distributional,
	title={A distributional framework for matched employer employee data},
	author={Bonhomme, St{\'e}phane and Lamadon, Thibaut and Manresa, Elena},
	journal={Econometrica},
	volume={87},
	number={3},
	pages={699--739},
	year={2019},
	publisher={Wiley Online Library}
}

@book{jiang2017asymptotic,
	title={Asymptotic analysis of mixed effects models: theory, applications, and open problems},
	author={Jiang, Jiming},
	year={2017},
	publisher={CRC press}
}

@article{bonhomme2022discretizing,
	title={Discretizing unobserved heterogeneity},
	author={Bonhomme, St{\'e}phane and Lamadon, Thibaut and Manresa, Elena},
	journal={Econometrica},
	volume={90},
	number={2},
	pages={625--643},
	year={2022},
	publisher={Wiley Online Library}
}

@article{bonhomme2022posterior,
	title={Posterior average effects},
	author={Bonhomme, St{\'e}phane and Weidner, Martin},
	journal={Journal of Business \& Economic Statistics},
	volume={40},
	number={4},
	pages={1849--1862},
	year={2022},
	publisher={Taylor \& Francis}
}

@article{card2013workplace,
	title={Workplace heterogeneity and the rise of West German wage inequality},
	author={Card, David and Heining, J{\"o}rg and Kline, Patrick},
	journal={The Quarterly journal of economics},
	volume={128},
	number={3},
	pages={967--1015},
	year={2013},
	publisher={MIT Press}
}

@article{song2019firming,
	title={Firming up inequality},
	author={Song, Jae and Price, David J and Guvenen, Fatih and Bloom, Nicholas and Von Wachter, Till},
	journal={The Quarterly journal of economics},
	volume={134},
	number={1},
	pages={1--50},
	year={2019},
	publisher={Oxford University Press}
}

@article{card2018firms,
	title={Firms and labor market inequality: Evidence and some theory},
	author={Card, David and Cardoso, Ana Rute and Heining, Joerg and Kline, Patrick},
	journal={Journal of Labor Economics},
	volume={36},
	number={S1},
	pages={S13--S70},
	year={2018},
	publisher={University of Chicago Press Chicago, IL}
}

@techreport{kane2008estimating,
	title={Estimating teacher impacts on student achievement: An experimental evaluation},
	author={Kane, Thomas J and Staiger, Douglas O},
	year={2008},
	institution={National Bureau of Economic Research}
}

@article{chetty2014measuring,
	title={Measuring the impacts of teachers I: Evaluating bias in teacher value-added estimates},
	author={Chetty, Raj and Friedman, John N and Rockoff, Jonah E},
	journal={American economic review},
	volume={104},
	number={9},
	pages={2593--2632},
	year={2014},
	publisher={American Economic Association 2014 Broadway, Suite 305, Nashville, TN 37203}
}

@book{efron2012large,
	title={Large-scale inference: empirical Bayes methods for estimation, testing, and prediction},
	author={Efron, Bradley},
	volume={1},
	year={2012},
	publisher={Cambridge University Press}
}

@article{laliberte2021long,
	title={Long-term contextual effects in education: Schools and neighborhoods},
	author={Lalibert{\'e}, Jean-William},
	journal={American Economic Journal: Economic Policy},
	volume={13},
	number={2},
	pages={336--377},
	year={2021},
	publisher={American Economic Association 2014 Broadway, Suite 305, Nashville, TN 37203-2425}
}

@article{gu2023invidious,
	title={Invidious comparisons: Ranking and selection as compound decisions},
	author={Gu, Jiaying and Koenker, Roger},
	journal={Econometrica},
	volume={91},
	number={1},
	pages={1--41},
	year={2023},
	publisher={Wiley Online Library}
}

@article{becker1973theory,
	title={A theory of marriage: Part I},
	author={Becker, Gary S},
	journal={Journal of Political economy},
	volume={81},
	number={4},
	pages={813--846},
	year={1973},
	publisher={The University of Chicago Press}
}

@article{eeckhout2011identifying,
	title={Identifying sorting—in theory},
	author={Eeckhout, Jan and Kircher, Philipp},
	journal={The Review of Economic Studies},
	volume={78},
	number={3},
	pages={872--906},
	year={2011},
	publisher={Oxford University Press}
}

@article{bonhomme2010generalized,
	title={Generalized non-parametric deconvolution with an application to earnings dynamics},
	author={Bonhomme, St{\'e}phane and Robin, Jean-Marc},
	journal={The Review of Economic Studies},
	volume={77},
	number={2},
	pages={491--533},
	year={2010},
	publisher={Wiley-Blackwell}
}

@article{efron2016empirical,
	title={Empirical Bayes deconvolution estimates},
	author={Efron, Bradley},
	journal={Biometrika},
	volume={103},
	number={1},
	pages={1--20},
	year={2016},
	publisher={Oxford University Press}
}

@article{bonhomme2015grouped,
	title={Grouped patterns of heterogeneity in panel data},
	author={Bonhomme, St{\'e}phane and Manresa, Elena},
	journal={Econometrica},
	volume={83},
	number={3},
	pages={1147--1184},
	year={2015},
	publisher={Wiley Online Library}
}

@article{bertrand2017field,
	title={Field experiments on discrimination},
	author={Bertrand, Marianne and Duflo, Esther},
	journal={Handbook of economic field experiments},
	volume={1},
	pages={309--393},
	year={2017},
	publisher={Elsevier}
}

@article{jochmans2019fixed,
	title={Fixed-effect regressions on network data},
	author={Jochmans, Koen and Weidner, Martin},
	journal={Econometrica},
	volume={87},
	number={5},
	pages={1543--1560},
	year={2019},
	publisher={Wiley Online Library}
}

@article{hull2018estimating,
	title={Estimating treatment effects in mover designs},
	author={Hull, Peter},
	journal={arXiv preprint arXiv:1804.06721},
	year={2018}
}

@article{arellano2009robust,
	title={Robust priors in nonlinear panel data models},
	author={Arellano, Manuel and Bonhomme, St{\'e}phane},
	journal={Econometrica},
	volume={77},
	number={2},
	pages={489--536},
	year={2009},
	publisher={Wiley Online Library}
}

@article{abowd2019modeling,
	title={Modeling endogenous mobility in earnings determination},
	author={Abowd, John M and McKinney, Kevin L and Schmutte, Ian M},
	journal={Journal of Business \& Economic Statistics},
	volume={37},
	number={3},
	pages={405--418},
	year={2019},
	publisher={Taylor \& Francis}
}

@article{finkelstein2016sources,
	title={Sources of geographic variation in health care: Evidence from patient migration},
	author={Finkelstein, Amy and Gentzkow, Matthew and Williams, Heidi},
	journal={The quarterly journal of economics},
	volume={131},
	number={4},
	pages={1681--1726},
	year={2016},
	publisher={MIT Press}
}

@article{amiti2018much,
	title={How much do idiosyncratic bank shocks affect investment? Evidence from matched bank-firm loan data},
	author={Amiti, Mary and Weinstein, David E},
	journal={Journal of Political Economy},
	volume={126},
	number={2},
	pages={525--587},
	year={2018},
	publisher={University of Chicago Press Chicago, IL}
}

@incollection{abowd2008econometric,
	title={Econometric analyses of linked employer--employee data},
	author={Abowd, John M and Kramarz, Francis and Woodcock, Simon},
	booktitle={The econometrics of panel data: Fundamentals and recent developments in theory and practice},
	pages={727--760},
	year={2008},
	publisher={Springer}
}

@article{heijmans1986consistent,
	title={Consistent maximum-likelihood estimation with dependent observations: the general (non-normal) case and the normal case},
	author={Heijmans, Risto DH and Magnus, Jan R},
	journal={Journal of Econometrics},
	volume={32},
	number={2},
	pages={253--285},
	year={1986},
	publisher={Elsevier}
}

@article{heijmans1986asymptotic,
	title={Asymptotic Normmality of Maximum Likelihood Estimators Obtained from Normally Distributed but Dependent Observations},
	author={Heijmans, Risto DH and Magnus, Jan R},
	journal={Econometric Theory},
	volume={2},
	number={3},
	pages={374--412},
	year={1986},
	publisher={Cambridge University Press}
}

@article{kotlarski1967characterizing,
	title={On characterizing the gamma and the normal distribution},
	author={Kotlarski, Ignacy},
	journal={Pacific Journal of Mathematics},
	volume={20},
	number={1},
	pages={69--76},
	year={1967},
	publisher={Mathematical Sciences Publishers}
}

@article{li1998nonparametric,
	title={Nonparametric estimation of the measurement error model using multiple indicators},
	author={Li, Tong and Vuong, Quang},
	journal={Journal of Multivariate Analysis},
	volume={65},
	number={2},
	pages={139--165},
	year={1998},
	publisher={Elsevier}
}

@article{lamadon2022imperfect,
	title={Imperfect competition, compensating differentials, and rent sharing in the US labor market},
	author={Lamadon, Thibaut and Mogstad, Magne and Setzler, Bradley},
	journal={American Economic Review},
	volume={112},
	number={1},
	pages={169--212},
	year={2022},
	publisher={American Economic Association 2014 Broadway, Suite 305, Nashville, TN 37203}
}

@article{sorkin2018ranking,
	title={Ranking firms using revealed preference},
	author={Sorkin, Isaac},
	journal={The quarterly journal of economics},
	volume={133},
	number={3},
	pages={1331--1393},
	year={2018},
	publisher={Oxford University Press}
}

@article{lentz2023anatomy,
	title={The Anatomy of Sorting—Evidence From Danish Data},
	author={Lentz, Rasmus and Piyapromdee, Suphanit and Robin, Jean-Marc},
	journal={Econometrica},
	volume={91},
	number={6},
	pages={2409--2455},
	year={2023},
	publisher={Wiley Online Library}
}

@inproceedings{james1961estimation,
	title={Estimation with quadratic loss},
	author={James, W., and C. Stein},
	booktitle={Proc. 4th Berkeley Symp. on Math. Statist. and Prob., 1961},
	year={1961}
}

@book{wasserman2006all,
	title={All of nonparametric statistics},
	author={Wasserman, Larry},
	year={2006},
	publisher={Springer Science \& Business Media}
}
}

\end{document}